\documentclass[aps,pra,amsfonts,amssymb,twocolumn,groupedaddress,showpacs]{revtex4}
\usepackage{graphics}
\usepackage{amsmath}

\newcommand{\Tr}{\mbox{Tr}}
\def\beq{\begin{equation}}
\def\eeq{\end{equation}}
\def\bea{\begin{eqnarray}}
\def\eea{\end{eqnarray}}

\begin{document}

\title{Qubit quantum channels: A characteristic function approach}

\author{Filippo Caruso and Vittorio Giovannetti}

\affiliation {NEST CNR-INFM \& Scuola Normale Superiore, Piazza
dei Cavalieri 7, I-56126 Pisa, Italy}

\begin{abstract}

A characterization of qubit quantum channels is introduced. In
analogy to what happens in the context of Bosonic channels we
exploit the possibility of representing the states of the system
in terms of characteristic function. The latter are functions of
non-commuting variables (Grassmann variables) and are defined in
terms of generalized displacement operators. In this context we
introduce the set of Gaussian channels and show that they share
similar properties with the corresponding Bosonic counterpart.

\end{abstract}


\maketitle

In quantum mechanics the transition from the initial state to the
final state of a system is described in terms of quantum
channels~\cite{nielsen}. At a mathematical level these are linear
maps operating on the set of bounded operators of the system,
which preserve the trace and (if any) the positivity of the
operators on which they act. Finally in order to represent a
``physical'' transformation, i.e., a transformation that could be
implemented in a lab, a quantum channel must also possess the
property of  complete positivity (i.e., the positivity of any
initial joint operator acting on the system plus an external
ancilla need to be preserved by the action of the map). An
impressive effort has been devoted in the last decades to study
the properties of quantum channels. Indeed they play a fundamental
role in many different branch of physics, specifically  in all
those sectors where one is interested in studying the decoherence
and noise effects.

In the context of quantum information theory emphasis is put on
characterizing the properties of quantum channels in terms of
their information capacities~\cite{SHOR,KEYL}. These figures of
merit are the quantum counterpart of the Shannon capacity of a
classical communication line~\cite{COVER}, which ``measure'' the
performances of the map in conveying classical or quantum
information. Even though impressive achievement has been obtained
in this field in the recent years, several open questions are
still under investigation --- we refer the reader to~\cite{LIST}
and references therein for details.

The majority of the results obtained so far relate to two specific
classes of channels, namely the qubit channels and the Bosonic
Gaussian channels. The former are completely positive
trace-preserving (CPT) transformations which act on the state of a
single two-level quantum system (qubit). Due to the small size of
the Hilbert space a simple parametrization of these channels has
been obtained~\cite{RUSKAI,ruskai2} while some additive
issues~\cite{DEVSHOR,KINGUNITAL,KOLDAN} and several classical and
quantum capacities~\cite{QC,DEVSHOR,wolf,KINGUNITAL,KINGDEP} have
successfully been solved (see also Ref.~\cite{KEYL} for a review).
Bosonic Gaussian channels~\cite{HW,HOLEVOBOOK}, on the contrary,
are a specific subclass of CPT maps acting on a continuous
variable system that preserve certain symmetries. These channels
include a variety of physical transformations that are of
fundamental interest in optics, including thermalization, loss and
squeezing. As in the qubit channel case, additivity
issues~\cite{GL,SEW} and
capacities~\cite{CAVES,LOSS,NOSTRO,NOSTRO1, WOLF1} have been
successfully solved for Bosonic Gaussian channels. Furthermore,
they allow for a compact parametrization
\cite{SEW,HOLEVONEW,NOSTRO1,WOLFLAST} in terms of the
characteristic function formalism~\cite{GLAUBEROPT,REV1,WMILBURN}.

In this paper we establish a parallelism among the qubit channels
and the Bosonic Gaussian channels by introducing for the former a
characteristic function representation. To do so we adapt the
formalism introduced by Cahill and Glauber in Ref.~\cite{CAHILL}
for representing the density operators of Fermions to the case of
two-level systems. In this context the channels  are represented
in terms of Green functions. Interestingly enough this allows us
to define a set of Gaussian  channels for qubit that share
analogous properties with their continuous variable counterpart.

The paper is organized as follows. In Sec.~\ref{s:uno} we briefly
review the characteristic function formalism for Bosonic (and
Fermionic) systems. In Sec.~\ref{s:due} we introduce the
displacement operator and characteristic function for a qubit. To
do so we introduce Grassmann variables and we use them to
generalize the definition of coherent states for finite
dimensional systems. We then present a Green function
representation for qubit channels (Sec.~\ref{s:tre}) and define
the set of qubit Gaussian channels (Sec.~\ref{s:sectioniv})
discussing their degradability properties. Conclusion and final
remarks are presented in Sec.~\ref{s:conclusion}. The paper
includes also a couple of technical Appendixes: namely, in
Appendix \ref{app1} we review some properties of Grassmann
calculus, while in Appendix \ref{s:degapp} we present a brief
excursus on quantum channel degradability.

\section{Characteristic function for Bosons}\label{s:uno}

In quantum optics a complete description of the state of a Bosonic
mode characterized by annihilation and creator operators $a$ and
$a^\dag$, can be obtained in terms of its coherent states
$|\mu\rangle$. These vectors possess various appealing properties.
Specifically, they minimize the uncertainty relations of any
couple of conjugate quadratures and they are eigenvectors of $a$.
Most importantly, coherent states form an over-complete continuous
set of vectors parametrized by a single  complex variable $\mu$.
This allows us to expand any other state of the system as a
superposition of the $|\mu\rangle$s with coefficients which define
quasi-probability density functions. Exploiting this and the fact
that the coherent states can be obtained by applying the
displacement operator $D(\mu) \equiv \exp[ \mu a^\dag - \mu^* a]$
to the vacuum, one can also use the latter as an over-complete
operator basis~\cite{WMILBURN,HOLEVOBOOK,REV1}. In particular,
given $\Theta$ any bounded operator of the system (e.g., a density
matrix $\rho$), we can write \bea \Theta = \int d^2\mu \;
\chi(\mu) \; D(- \mu) \label{invcar} \;, \eea where $d^2 \mu
\equiv d\text{Re}(\mu) d\text{Im}({\mu})$ and where \bea \chi( \mu
)\equiv \Tr [\Theta D(\mu)] \label{caratt} \;. \eea
Equation~(\ref{caratt}) defines the {\em characteristic} function
of the operator $\Theta$. This is a complex function of the
variables $\mu$ and $\mu^*$  which provides us with a faithful
description of the original operator thanks to the
``orthogonality'' relation \bea \Tr [ D(\mu) D( - \nu) ] =
\delta^{(2)} (\mu - \nu) \;, \eea with $\delta^{(2)}(\mu -\nu)$
being the Dirac delta in the complex plane. To  represent a
density matrix $\rho$, the function~(\ref{caratt})  needs to
possess certain properties~\cite{HOLEVOBOOK,REV1} including being
continuously differentiable in $\mu$ and $\mu^*$ and verifying
$\chi(0)=1$. Within the characteristic function description,
Gaussian states are defined as those $\rho$ whose $\chi(\mu)$ is a
Gaussian function of the complex parameter $\mu$ (examples are
thermal, coherent, and squeezed states).

Consider now the action of a linear super-operator $\cal N$ which
transforms $\Theta$ into $\Theta^\prime = {\cal N}(\Theta)$.
Equation~(\ref{invcar}) allows us to represent this mapping in
terms of a linear transformation of ${\chi}(\mu)$. Indeed  the
characteristic function of the output operator $\Theta^\prime$ is
\bea \chi^\prime( \mu ) &=& \Tr [\Theta^\prime D(\mu)] = \Tr[
\Theta {\cal N}_H[ D(\mu) ]]
\nonumber \\
&=&\int d^2 \nu \; \chi(\nu)  \; G(\nu,\mu) \label{carattout1} \;,
\eea with \bea G(\nu, \mu) &\equiv& \Tr[ D(-\nu) {\cal N}_H\Big(
D(\mu)\Big) ]
\label{green0} \\
&=& \Tr[ {\cal N }\Big( D(-\nu) \Big)  D(\mu) ] \;. \label{green}
\eea In these expressions ${\cal N}_H$ is the dual of ${\cal N}$
which describes the channel in the Heisenberg picture and which is
defined by the identity
\begin{eqnarray}
\mbox{Tr} [ \Theta_1 {\cal N}_H (\Theta_2) ] =\mbox{Tr} [{\cal N} (\Theta_1) \Theta_2]
\label{dual}\;,
\end{eqnarray}
for all $\Theta_{1,2}$ (see, for instance, \cite{NOSTRO1}). We
call Eq. (\ref{green0})  the Green function of ${\cal N}$:
according to the previous definitions it provides us with a
complete characterization of the channel.

A special subset of CPT maps for Bosonic systems is the set of
Gaussian channels~\cite{HW}. These are characterized by Green
functions~(\ref{green}) of the form \bea
 G(\nu, \mu) &=&
\delta^{(2)} ( \nu - v \mu - w \mu^*)\nonumber \\
&&\times  \exp \left[ - \frac{1}{2}  (\mu^* ,\ - \mu) \Gamma\binom
{\mu} {-\mu^*} \right] \;,\label{GAUSBOS}
 \eea
with $\Gamma$ being a real symmetric positive matrix (i.e.,
covariance matrix) and $v$ and $w$ being complex numbers ---
rigorously speaking, Eq.~(\ref{GAUSBOS}) defines one-mode Bosonic
Gaussian channels. As can be directly verified from
Eq.~(\ref{carattout1}) such maps have the peculiar property of
transforming input Gaussian states into output Gaussian states. An
interesting fact about these channels is that, except for the
additive classical noise channel~\cite{HOLEVONEW,NOSTRO1}, they
admit a physical representation~\cite{NOSTRO1} in terms of a
single mode environment originally prepared in a Gaussian state.
Indeed, the exceptional role of the additive classical noise
channel corresponds to the fact that any one-mode Bosonic Gaussian
channel can be represented as a unitary coupling with a
single-mode environment plus an additive classical noise. Within
such representation (without additive classical noise) one can
show that the Bosonic Gaussian channels~(\ref{GAUSBOS}) are either
anti-degradable or weakly degradable~\cite{NOSTRO}. Moreover, in
the case in which the single-mode representation is of Stinespring
form (that is, if the environment state is pure) the channel is
then anti-degradable or degradable in the sense of
Ref.~\cite{DEVSHOR} (for the sake of completeness explicit
definitions of these properties are given in Appendix
\ref{s:degapp}).

\subsection{Characteristic function for Fermions}\label{s:charafer}

The characteristic function formalism presented in the previous
section can be generalized to describe Fermionic systems
too~\cite{CAHILL}. The main difference in this case is related to
the fact that now the complex variables $\mu$ and $\mu^*$ are
replaced by a couple of conjugate Grassmann variables $\xi$ and
$\xi^*$~\cite{GRASS} whose properties are reviewed in
Appendix~\ref{app1}. This is intrinsically related  with the fact
that the annihilation and creation operators of a  Fermion obey
anti-commutation rules instead of commutation
rules~\cite{schwinger}. We will not review the analysis of
Ref.~\cite{CAHILL} since in the next section, when discussing the
qubit case, we will rederive most of the results obtained in the
Fermionic case.

\section{Representation of a qubit}\label{s:due}

Various proposals for defining a (discrete) phase space for finite
dimensional systems have been discussed so far by introducing
generalized position and momentum operators (see, for instance,
Ref.~\cite{WOOT} and references therein). Here we will not follow
this line: instead we invoke the analogies between a qubit and a
single Fermionic mode to adapt the results of Ref.~\cite{CAHILL}.
A similar approach was developed in Ref.~\cite{ANAS} to solve
non-Markovian master equations of a two-level atom interacting
with an external field.

The starting point of our analysis is to observe that the lowering
and raising operators of the qubit [i.e., $\sigma_+  \equiv
|1\rangle\langle 0| $ and $\sigma_- \equiv (\sigma_+)^\dag$]
satisfy anti-commutation rules similar to those of a Fermionic
mode, i.e., \bea \{ \sigma_-, \sigma_+ \} &=& |0\rangle\langle 0|
+ |1\rangle\langle 1| \equiv
\openone \; , \nonumber \\
\{ \sigma_-, \sigma_- \} &=& \{ \sigma_+, \sigma_+ \} = 0 \;.
\label{antiqu} \eea Identifying the qubit  state $|0\rangle$ with
the Fermionic vacuum we can therefore treat $\sigma_+$ and
$\sigma_-$ as Fermionic creation and annihilation operators,
respectively. Following~\cite{CAHILL} we introduce then a couple
of conjugate Grassmann variables $\xi$ and $\xi^*$ (see Appendix
\ref{app1}) and impose standard anti-correlation with the
annihilation and creator operators of the system, i.e., \bea \{
\xi ,\sigma_{\pm} \} = \{ \xi^*, \sigma_{\pm} \} = 0 \;.
\label{commop} \eea It is worth noticing that this implies that
the projectors
 $|0\rangle\langle 0| = \sigma_-\sigma_+$ and
$|1\rangle\langle 1| = \sigma_+\sigma_-$ as well as the
Pauli matrix $\sigma_z \equiv |0\rangle\langle 0| - |1\rangle\langle 1|$
commute with $\xi$ and $\xi^*$.

In the following we will also require that
 \bea
\xi \; | j \rangle &=& (-1)^j | j\rangle \; \xi \nonumber  \\
\xi^* \; | j \rangle &=& (-1)^j | j\rangle \; \xi^* \;,
\label{vec} \eea for $j=0,1$. This is not strictly necessary but
it is consistent with Eq.~(\ref{commop}) and allows us to simplify
the calculations. For instance, given any collection of qubit
operators $\Theta_1$, $\Theta_2$, $\cdots$, $\Theta_{n+1}$ and the
Grassmann numbers $\xi_1$, $\xi_2$, $\cdots$, $\xi_n$  we can use
Eq.~(\ref{vec}) to verify that the following relation applies \bea
&&\Tr [ \Theta_1 \xi_1 \Theta_2 \xi_2 \cdots \Theta_n \xi_n
\Theta_{n+1} ] \label{traccia}\\ \nonumber && \qquad = \xi_1 \xi_2
\cdots \xi_n \; \Tr[ \Theta_1 \sigma_z \Theta_2 \sigma_z \cdots
\Theta_n\sigma_z\Theta_{n+1} ] \eea (an analogous expression holds
also when replacing all, or part of, the $\xi_i$s with their
complex conjugates --- more details about the trace can be found
in Appendix~\ref{app2}).

The above definitions give us  the possibility of operating with
``hybrid'' mathematical objects obtained by multiplying Grassmann
variables and qubit operators. In this context we find it useful
to define a generalized adjoint operation for these hybrid
operators by arbitrarily imposing the conditions \bea
&&\left(\Theta_1 \xi_1 \Theta_2 \xi_2 \cdots \Theta_n \xi_n
\Theta_{n+1} \right)^\dag \nonumber \\
&& \qquad =
\Theta_{n+1}^\dag \; \xi_n^* \Theta_n^\dag \cdots
 \; \xi_2^*  \; \Theta_2^\dag \; \xi_1^* \;
\Theta_1^\dag   \label{adjont} \;, \eea with $\xi_{i}$  and
$\Theta_i$ as in Eq.~(\ref{traccia}).

\subsection{Qubit characteristic function}

Qubit displacement operators can now be defined in analogy
with~\cite{CAHILL} as
\begin{eqnarray}
 D (\xi) &\equiv& \exp \left(
\sigma_+ \xi - \xi^* \sigma_- \right) \label{DPQ} \\
&=& \openone + \sigma_+ \xi - \xi^* \sigma_-  - \sigma_z\xi^* \xi
/2 \; , \nonumber
\end{eqnarray}
where in the second line we used Eq.~(\ref{fexp}). As in the
Bosonic case they satisfy the identity $D^\dag(\xi) = D(-\xi)$.
Moreover the application of $D(\xi)$ to the vacuum originates
eigenvectors of the annihilation operator of the system
($\sigma_-$). These are the coherent states of our qubit, i.e.,
\begin{eqnarray}
|\xi\rangle =  D (\xi) |0\rangle = \left(1 - \frac{\xi^*\xi}{2}\right) |0\rangle
- \xi |1\rangle
\label{coqu}\;,
\end{eqnarray}
whose norm is unity. These vectors are eigenvectors of $\sigma_-$
in Grassmann sense (i.e., their eigenvalues are Grassmann
variables; see Ref. \cite{CAHILL} for details).

What is interesting for us  is the fact that $D(\xi)$ can be used
to define  a characteristic function for the operators of the
system as in Eq.~(\ref{caratt}), i.e., \bea \chi( \xi )\equiv \Tr
[\Theta D(\xi)] \label{caratt1} \;. \eea

In particular, consider a $\Theta$ which is characterized by the
matrix
\bea \Theta\equiv \left(\begin{array}{cc}
\theta_{00} & \theta_{01} \\
\theta_{10} & \theta_{11} \end{array} \right) \;,\label{qubit}
\eea when expressed in the computational basis $\{ |0\rangle$,
$|1\rangle \}$. In this case using the anti-commutation rules of
Eq.~(\ref{commop})
 and the identity (\ref{traccia})
we get \bea \chi(\xi) = \Tr[\Theta] + (\theta_{00} -\theta_{11} )
\frac{\xi \xi^*}{2}
 + \theta_{01} \xi - \theta_{10} \xi^*. \label{QQQ}
\eea

It is worth noticing that with respect to the analysis of
Ref.~\cite{CAHILL} the characteristic functions analyzed here
contain an extra term which is linear in $\xi$ and $\xi^*$. Indeed
in the Fermionic case analyzed by Cahill and Glauber the only
allowed physical states are classical mixtures of $|0\rangle
\langle 0|$ and $|1\rangle\langle 1|$ (this follows from the
requirement of invariance under $2\pi$ rotation with respect to an
arbitrary axis). Consequently the off-diagonal terms associated
with $\theta_{01}$ and $\theta_{10}$ do not need to be considered.
When analyzing qubit systems, instead, quantum superpositions
among $|0\rangle$ and $|1\rangle$ are allowed and we need to
include also the linear contributions.

As in the Bosonic case, Eq.~(\ref{caratt1}) can be inverted. In
this case, however, Eq.~(\ref{invcar}) is replaced by \bea \Theta
= \int d^2 \xi \; \chi(\xi) \; \tilde{E} (-\xi) \label{inv1} \;,
\eea with $\tilde{E}(\xi) \neq D(\xi)$ defined by
\begin{eqnarray}
 \tilde{E}(\xi) &\equiv&
\sigma_z -\xi^* \xi /2 +  \sigma_+ \xi - \xi^* \sigma_-
\; . \label{ETILDE}
\end{eqnarray}
The easiest way to verify this is by direct substitution of
Eqs.~(\ref{QQQ}) and (\ref{ETILDE}) into Eq.~(\ref{inv1}) and by
employing the integration rules~(\ref{integral}).

\subsubsection{Density operators}

To represent a density operator \bea
 \rho \equiv
\left(\begin{array}{cc}
p  & \gamma \\
\gamma^* & 1-p  \end{array} \right) \label{rho}
 \eea
the characteristic function needs to satisfy certain physical
requirements. First of all,  the Hermitianity of $\rho$ and the
normalization condition $\Tr[\rho]=1$ imply, respectively, \bea
\chi(\xi) &=& \left[\chi(-\xi)\right]^* \label{hermi} \;, \\
\chi(0) &=& 1\;, \label{norm} \eea where complex conjugation is
defined as in Eq.~(\ref{complexconj}) [to verify this simply use
Eq.~(\ref{QQQ}) with $\Theta =\rho$]. The positivity of $\rho$
imposes, instead, the following inequality to hold
 \beq
 \left|\int d^2 \xi \ \chi(\xi) \xi\right|^2 + \left[\int d^2 \xi \ \chi(\xi)\right]^2
\leqslant \frac{1}{4} \;. \label{pos}
 \eeq
This follows from the positivity condition $|\gamma|^2 \leqslant p
(1-p)$ and by the identity \bea
 \gamma &=& \int d^2 \xi \; \chi(\xi) \; \xi^* \;,\nonumber \\
  p  &=& \int d^2 \xi\ \chi(\xi) + 1/2 \nonumber \;.
\eea Using similar arguments one can verify that
Eqs.~(\ref{hermi})-(\ref{pos}) are also sufficient conditions for
$\chi(\xi)$ being a characteristic function of a density operator
$\rho$.

\section{Green function representation of a qubit channel}\label{s:tre}

Let us now consider the effect of a qubit quantum channel ${\cal
N}$ acting on the operator $\Theta$ of the system. As in the
Bosonic case we would like to derive its Green function
representation~(\ref{green}). To do so we first evaluate the
characteristic function $\chi^\prime (\xi)$ associated with $A
\Theta B$ with $A$ and $B$ being arbitrary qubit operators. This
is \bea
\chi^\prime (\xi ) &=& \Tr [ A\Theta B D(\xi)] \nonumber \\
&=& \int d^2\zeta  \;\Tr[ A \chi(\zeta) \tilde{E}(-\zeta) B
D(\xi)] \;, \label{ffef} \eea where we used Eq.~(\ref{inv1}) with
$\chi(\xi)$ being
 the characteristic
function of $\Theta$ (from now on $\zeta$ and $\xi$ should be considered entries of the
same Grassmann set).
Our goal is to find a function
$G(\zeta,\xi)$ which gives \bea \chi^\prime (\xi ) &=& \int
d^2\zeta  \; \chi(\zeta) \; G(\zeta,\xi)  \;, \label{goal} \eea
 for all $\chi(\xi)$. Notice that if
$\xi$ were a commuting variable (e.g., a complex variable) the
problem could be solved by simply moving $\chi(\xi)$ out of the
trace operation of Eq.~(\ref{ffef}) yielding $G(\zeta,\xi) = \Tr[
A \tilde{E}(-\zeta)  B D(\xi)] $. In the case under consideration,
however, the situation is complicated  by the fact that for moving
out of trace  the variables $\xi$ or $\xi^*$ we need to insert
$\sigma_z$s as in Eq.~(\ref{traccia}). Taking into account this
fact, the solution becomes \bea G(\zeta,\xi) &=& \Tr [ A \sigma_z
D(-\zeta) B D(\xi)] \;, \eea as can be easily verified by direct
integration of the Eqs.~(\ref{ffef}) and (\ref{goal}) for the most
general characteristic function~(\ref{QQQ}).

The Green function~(\ref{goal}) associated with a CPT map ${\cal
N}$ can then be obtained by using an operator sum
representation~\cite{nielsen,KEYL} of such channel and exploiting
the linearity of the trace. Indeed, writing ${\cal N} (\Theta) =
\sum_k M_k \Theta M_k^\dag$ with $\{ M_k\}_k$ being
 Kraus operators of ${\cal N}$, we get
\bea G(\zeta,\xi) &=& \sum_k \Tr [ M_k \sigma_z D(-\zeta) M_k^\dag
D(\xi)]
\nonumber \\
&=& \Tr \Big[ {\cal N} \Big( \sigma_z D(-\zeta) \Big ) D(\xi)\Big]
\;. \label{GREEN} \eea Using Eq.~(\ref{QUIE}) this can also be
written as \bea G(\zeta,\xi) &=& \Tr \Big[  \sigma_z D(-\zeta)
{\cal N}_H \Big(D(\xi)\Big) \Big] \;, \label{GREEN0} \eea with
${\cal N}_H$ being the Heisenberg representation of the map ${\cal
N}$ defined in Eq.~(\ref{dual}). Equation~(\ref{GREEN0}) shows
that, as in the Bosonic case, a complete description of the
channel is obtained by applying the dual map to the displacement
operator
--- see Eq.~(\ref{green0}).
Exploiting the normalization condition $\sum_k M_k^\dag M_k =\openone$ we note
 that for $\xi=0$ the above expression yields
\begin{eqnarray}
G(\zeta,0)=\Tr [ \sigma_z D(-\zeta) ]= \zeta \zeta^*
\;,
\end{eqnarray}
which corresponds
 to the  Grassmann delta function $\delta^{(2)}(\zeta)$ defined in  Eq.~(\ref{deltaf}),
in agreement with the requirement of channel being trace preserving
--- see Eqs.~(\ref{QQQ}) and (\ref{goal}).

Finally, let ${\cal N}_1$ and ${\cal N}_2$ be two different qubit channels
with Green functions $G_1(\zeta,\xi)$ and $G_2(\zeta,\xi)$,
respectively. From the definition~(\ref{goal}) we then find that the
 Green function $G_{12}(\zeta,\xi)$ of the
composite map ${\cal N}_2 \circ {\cal N}_1$ in which  we first
operate with ${\cal N}_1$ and then with ${\cal N}_2$, can be
expressed in terms of the following Grassmann convolution integral
 \bea
 G_{12}(\zeta,\xi)= \int d^2 \xi^\prime  \;
G_1(\zeta,\xi^\prime) \; G_2(\xi^\prime,\xi) \label{compos},
 \eea
with $\zeta$, $\xi$, and $\xi^\prime$ Grassmann numbers.

\subsection{Examples and canonical forms} \label{s:examples}

As a particular case of Green function consider the identity map
${\cal I}$ which leaves all operators invariant, i.e., ${\cal
I}(\Theta) = \Theta$. According to our definition we get \bea
G(\zeta,\xi) =\Tr [  \sigma_z D(-\zeta)
 D(\xi)]
= (\zeta -\xi) (\zeta^* -\xi^*) \;, \label{GREENid}
 \eea
which, as expected, corresponds to the  delta $\delta^{(2)}
(\zeta-\xi)$ of Eq.~(\ref{deltaf}). More generally from
Ref.~\cite{RUSKAI} we know that the most generic qubit quantum
channel ${\cal N}$ implements the following transformation,
 \bea
{\cal N}(\rho)={\cal N}\left( \frac{\openone + {\vec{r}} {\mathbf
\cdot \vec{\sigma}}}{2} \right) = \frac{\openone +   ({\vec{t}}
+  T {\vec{r}}) \cdot \vec{\sigma} }{2} \;, \label{canono}
 \eea
where ${\vec{t}} = ({t}_1, {t}_2, {t}_3)$ is a real vector,
$\vec{\sigma}=\{\sigma_1,\sigma_2,\sigma_3\}$ is a vector
containing the Pauli matrices, ${\vec{r}}$ is the Bloch vector
describing the input state, and ${T}$ is a real $3 \times 3$
matrix. In Ref. \cite{RUSKAI} it is shown that $T$ can be reduced,
via changes of basis in $\mathbb C^2$ (i.e., via proper rotations
of the input and output states), to the diagonal (canonical) form
$T=diag(\lambda_1, \ \lambda_2, \ \lambda_3)$, with the real
coefficients $\lambda_{1,2,3}$ and $t_{1,2,3}$ that need to
satisfy certain conditions~\cite{RUSKAI,ruskai2} to guarantee the
complete positivity of the map. In the Green function language
such canonical form corresponds to have
 \bea
&& G(\zeta,\xi) = \label{green.qubit}
\\ \nonumber
&& \qquad  \delta^{(2)}
 \left(\zeta-\frac{\lambda_2+\lambda_1}{2} \xi
 - \frac{\lambda_2-\lambda_1}{2} \xi^* \right) \exp\left[-\frac{t_3}{2} \xi^* \xi \right]
 \nonumber \\
&&\quad +(\lambda_3-\lambda_1 \lambda_2)\xi \xi^*+\frac{t_1-i
t_2}{2} \zeta \zeta^*\xi -\frac{t_1+i t_2}{2}\zeta \zeta^*\xi^*
\;.  \nonumber
 \eea

\section{Gaussian channels for qubits} \label{s:sectioniv}

In analogy with the Bosonic case, in this section we introduce the
definition of qubit Gaussian channels. We start noticing that in
order to define these channels it does not make sense to focus on
maps which transform Gaussian characteristic functions into
Gaussian characteristic functions. Indeed, thanks to
Eq.~(\ref{fexp}), {\em all} characteristic functions of a qubit
can be written in a Gaussian form~\cite{NOTAG}. Therefore
following Eq. (\ref{GAUSBOS}) we say that a qubit map is Gaussian
if its Green function has the form
 \bea
G(\zeta,\xi) = \delta^{(2)} ( \zeta - a \xi - b \xi^*) \; \exp[- c
\xi^* \xi] \label{guass.cond} \;,
 \eea
with $a$ and $b$ complex and $c$ real~\cite{NOTA1} numbers,
respectively, and with the exponential defined as in
Eq.~(\ref{fexp}). A trivial example is provided by the identity
map ${\cal I}$ whose Green function~(\ref{GREENid}) is of the
form~(\ref{guass.cond}) for $b=c=0$ and $a=1$.

Generic mixtures of Gaussian channels do not necessarily have the
form~(\ref{guass.cond}). Therefore the set of Gaussian channels is
not convex. However, it has semi-group structure with respect to
the channel composition rule $\circ$. Indeed, given two Gaussian
channels ${\cal N}_1$ and ${\cal N}_2$ characterized by parameters
$(a_1, b_1, c_1)$ and $(a_2,b_2,c_2)$, respectively, from
Eq.~(\ref{compos}) it is easy to verify that the Green function of
${\cal N}_2 \circ {\cal N}_1$ is again of the form
(\ref{guass.cond}) with
\begin{eqnarray}
a &=& a_1 a_2 + b_1 b_2^* \;, \nonumber \\
b &=& a_1 b_2 + b_1 a_2^*  \;, \nonumber \\
c &=& c_1 ( |a_2|^2 - |b_2|^2) + c_2\;. \label{COMPOGU}
\end{eqnarray}
Both the semi-group property and the non-convexity property hold
also in the Bosonic case.

\subsection{Canonical form for Gaussian channels}\label{s:canogaus}
From Eq.~(\ref{green.qubit}) it is easy to verify that
within the parametrization~\cite{RUSKAI,ruskai2}
we can get Gaussian maps~(\ref{guass.cond})
by choosing
\begin{eqnarray}
\lambda_3 &=& \lambda_1 \lambda_2 \;, \qquad \\
t_1&=&t_2=0  \;.
\label{semplici}
\end{eqnarray}
This in fact yields Gaussian Green functions with $a =
({\lambda_2+\lambda_1})/{2}$, $b= ({\lambda_2-\lambda_1})/{2}$ and
$c = t_3/2$. We can then use~\cite{ruskai2} to show that the
corresponding transformation is CPT if and only if the following
inequalities hold,
 \bea
\left\{ \begin{array}{ccll}
 |\lambda_k|  &\leqslant& 1 &   \mbox{for $k=1$, $2$;}  \\
 \\
 |t_3| &\leqslant& \sqrt{(1-\lambda_1^2) (1-\lambda_2^2)} \;.&
\end{array}
\right. \label{cond3}  \eea

This allows us to parametrize the whole set of Gaussian channels
in terms of three real parameters only. First of all,   as in
Refs.~\cite{wolf,ruskai2}, we can use a trigonometric
parametrization to express $\lambda_{1,2}$ in terms of the angles
$\theta$, $\phi$ in $[0,2\pi[$ as follows
 \bea
 \lambda_1= \cos(\theta - \phi)\;, \quad \qquad  \lambda_2=\cos(\theta +
 \phi)\;. \label{para.com}
 \eea
Then we can parametrize $t_3$ by introducing the positive quantity
$q\in [0,1]$ to write
  \bea
 t_3= (2 q-1) \frac{ \cos(2\theta) - \cos(2\phi)}{2}\;. \label{para.com0}
 \eea
Replacing all this into Eq.~(\ref{green.qubit}) yields the
following canonical form for the Green function of a qubit
Gaussian channel, i.e.,
 \bea
 &&G(\zeta,\xi) = \delta^{(2)} \left(\zeta - \xi \cos \theta \cos \phi +\xi^*
\sin \theta \sin
\phi \right) \nonumber \\
&&\quad \times \exp \left[ (2q-1) \;\frac{\cos (2 \theta) - \cos
(2 \phi)}{4} \xi \xi^*
   \right]  \label{gfqubitmixed} \;.
 \eea
We will see that the maps of this form have the peculiar property
that they can always be described in terms of a unitary
interaction of the form~(\ref{HGCuno}) with a {\em single} (not
necessarily pure) qubit environment. For this reason we call them
``qubit-qubit'' channels. It is worth stressing that once again a
similar property holds for the Bosonic case: there (almost) all
the one-mode Bosonic Gaussian maps are in fact describable in
terms of a single mode environment~\cite{NOSTRO,NOSTRO1}.

\subsection{Qubit-qubit maps: Pure environment case} \label{ququ}

An important subclass of the qubit-qubit channels of
Eq.~(\ref{gfqubitmixed}) is obtained for $q=1$ and $\theta$ and
$\phi$ generic, i.e.,
 \bea
 &&G(\zeta,\xi) = \delta^{(2)} \left(\zeta - \xi \cos \theta \cos \phi +\xi^*
\sin \theta \sin
\phi \right) \nonumber \\
&&\quad \times \exp \left[\frac{\cos (2 \theta) - \cos
(2 \phi)}{4} \xi \xi^*
   \right]  \label{gfqubitpure} \;.
 \eea
According to Eq.~(\ref{cond3}) this corresponds to having $|t_3| =
\sqrt{(1 -\lambda_1^2) (1-\lambda_2^2)}$. As shown in
Ref.~\cite{ruskai2} any CPT map which can be described in terms of
an interaction with a single qubit environment originally prepared
in a pure state can be expressed in this form by proper unitary
rotation of the input and the output state. This implies  that the
maps~(\ref{gfqubitpure}) admit a Stinespring
dilation~(\ref{HGCuno}) with a two-dimensional (qubit) environment
$E$. Without loss of generality, we can assume an initial state of
the environment of the form $\rho_E \equiv |0\rangle_E\langle 0|$.
Following Ref. \cite{nielsen}, one can then choose the unitary
coupling $U$ to have the following block structure
\begin{eqnarray}
U = \left( \begin{array}{cc}
{[A_0]} & {[-\sigma_x A_1 \sigma_x]} \\
{[A_1]} & {[\sigma_x A_0 \sigma_x]}
\end{array} \right) \;, \label{unitary}
\end{eqnarray}
with
\begin{equation}
\label{kraus:qq}
A_0=\left(
\begin{array}{cc}
  \cos\theta & 0 \\
  0 & \cos\phi \\
\end{array}
\right),\quad
A_1=\left(
\begin{array}{cc}
  0 & \sin\phi \\
  \sin\theta & 0 \\
\end{array}
\right)\;,
\end{equation}
being a Kraus set for the channel [the matrix~(\ref{unitary}) is
expressed in the basis $\{ |00\rangle, |10\rangle,
|01\rangle,|11\rangle\}$ with $|jk\rangle \equiv |j\rangle \otimes
|k\rangle_E$ for $j,k=0,1$].

The complementary channel $\tilde{\cal N}$~\cite{DEVSHOR,HOLEVOREP,KING}
 can now be computed as in Eq.~(\ref{duedue}). Since
it represents a qubit channel --- it connects two two-dimensional Hilbert spaces
(the input Hilbert space with the environmental one) ---
we can use Eq.~(\ref{GREEN}) to evaluate its  Green function obtaining
 \bea
 \tilde{G}(\zeta,\xi) &=& \delta^{(2)} \left(\zeta - \xi \cos \theta \sin \phi +\xi^*
\sin \theta \cos
\phi \right) \nonumber \\
&& \ \ \ \times  \exp\left[\frac{\cos (2 \theta) + \cos (2
\phi)}{4} \xi \xi^*
   \right] \;. \label{compl}
 \eea
It is still of the (pure-environment qubit-qubit) Gaussian
form~(\ref{gfqubitpure}) and can be expressed in terms of the
original Green function $G(\zeta,\xi)$ of ${\cal N}$ by simply
shifting $\phi$ by $-\pi/2$ and by changing sign to $\theta$,
i.e.,
 \bea
 \tilde{G}(\zeta,\xi) &=& G(\zeta, \xi)\Big|_{\begin{subarray}{l}
\theta\rightarrow -\theta\\
\phi\rightarrow \phi -\pi/2
\end{subarray}} \label{gtildeg}\;.
 \eea
In Ref.~\cite{wolf} it has been shown that  qubit-qubit channels
with pure environment are degradable for
$\cos(2\theta)/\cos(2\phi) \geqslant 0$, and anti-degradable
otherwise. Here we will rederive this same result in the Green
function formalism as a consequence of the Gaussianity of these
maps, pointing out an interesting parallelism with their Bosonic
counterpart.

In analogy with~\cite{NOSTRO,NOSTRO1} we look for
the intermediate map ${\cal T}$ that should connect ${\cal N}$
with $\tilde{\cal  N}$, in the class of qubit-qubit channels (with
pure environment). Rewriting the degradability
condition~(\ref{degrado}) in terms of the compositions
rules~(\ref{compos}) we can then recast the problem as follows
 \bea
 \tilde{G}(\zeta,\xi)= \int d^2 \xi^\prime  \;
G(\zeta,\xi^\prime) \;
{G}_x(\xi^\prime,\xi) \;,
\label{degrad}
 \eea
where ${G}_x(\zeta, \xi)$ is the Green
function~(\ref{gfqubitpure}) of the map ${\cal T}$ characterized
by the parameters $\theta_x$ and $\phi_x$. By using
Eq.~(\ref{COMPOGU}) we find that, for $\cos(2\theta)/\cos(2\phi)
\geqslant 0$, $\theta_x, \phi_x$ do exist such that
Eq.~(\ref{degrad}) is satisfied. Specifically such parameters are
defined by the relations,
\begin{eqnarray}
\cos(2 \theta_x) &=& \frac{\cos(2 \theta)-\cos(2 \phi)+2\cos(2
\theta)\cos(2 \phi)}{\cos(2 \theta)+\cos(2 \phi)} \; , \nonumber \\
\cos(2 \phi_x) &=& \frac{\cos(2 \theta)-\cos(2 \phi) - 2\cos(2
\theta)\cos(2 \phi)}{\cos(2 \theta)+\cos(2 \phi)} \nonumber
\;. \\ \label{condiz2}
\end{eqnarray}

The case $\cos(2\theta)/\cos(2\phi) \leqslant 0$ can be treated
analogously to show that the corresponding channels are
anti-degradable. In fact, in the Green function formalism the
anti-degradability condition~(\ref{antidoto}) becomes
 \bea
 G(\zeta,\xi)= \int d^2 \xi^\prime  \;
\tilde{G}(\zeta,\xi^\prime) \;
\bar{G}_x(\xi^\prime,\xi)  \label{antidegrad} \;,
 \eea
where $\bar{G}_x(\zeta,\xi)$ is the Green function  of the
connecting map $\overline{\cal T}$. We find that for
$\cos(2\theta)/\cos(2\phi) \leqslant 0$, Eq.~(\ref{antidegrad}) is
satisfied by choosing $\bar{G}_x(\zeta,\xi)$ in the subclass of
qubit-qubit channels with pure environment -- i.e.,
Eq.~(\ref{gfqubitpure}) --
 with $\theta_x$ and
$\phi_x$ determined by the expressions (\ref{condiz2}) after
replacing $(\theta,\phi)$ with  $(-\theta,\phi-\pi/2)$.

More directly this result can be established by using the
correspondence~(\ref{gtildeg}) and the fact that the complementary
channels of degradable maps are anti-degradable --- see
Appendix~\ref{s:degapp}. Consider, in fact, a (pure environment)
qubit-qubit channel ${\cal N}$ with $\cos(2\theta)/\cos(2\phi)
\leqslant 0$. According to Eq.~(\ref{gtildeg}) we know that its
complementary $\tilde{\cal N}$ is still a (pure environment)
qubit-qubit channel characterized by the parameters
$(\theta^\prime,\phi^\prime) = (-\theta, \phi-\pi/2)$. Now it is
easy to verify that $\cos(2\theta^\prime)/\cos(2\phi^\prime)=
-\cos(2\theta)/\cos(2\phi)\geqslant 0$. Therefore from
Eqs.~(\ref{degrad}) and (\ref{condiz2}) we can conclude that
$\tilde{\cal N}$ is degradable while ${\cal N}$ is
anti-degradable.

Note that, in the special case $\cos (2 \theta)=\cos (2 \phi)=0$,
both the degradability relations are satisfied. Therefore in this
case the qubit-qubit channels with pure environment are both
degradable and anti-degradable, with null quantum capacity.

\subsection{Qubit-qubit maps: Mixed environment case}

Now let us consider the Gaussian channels~(\ref{gfqubitmixed}) for
$q\neq 1$. They can be represented in terms of a
physical representation~(\ref{HGCuno}) with $U$ as in Eq.~(\ref{unitary}) and with
$E$ being a single qubit
environment initially prepared in the mixed state,
\begin{eqnarray}
\rho_E \equiv q | 0 \rangle_E
\langle 0 | + (1-q) | 1 \rangle_E \langle 1 |
\label{envi} \;.
\end{eqnarray}
To verify this, we observe that with the above prescriptions
Eq.~(\ref{HGCuno}) gives
\begin{eqnarray}
{\cal N}(\rho) &=& \mbox{Tr}_E[ U \left\{\rho \otimes \left[q | 0
\rangle_E\langle 0|+(1-q) | 1 \rangle_E \langle 1 | \right]
\right\}
U^\dag] \nonumber \\
&=& q {\cal N}_0 (\rho) + (1-q) {\cal N}_1(\rho)
\;, \label{eeew}
\end{eqnarray}
with ${\cal N}_0 \equiv \mbox{Tr}_E [ U (\rho \otimes
|0\rangle_E\langle 0|) U^\dag]$ being the (pure environment)
qubit-qubit channel of Sec.~\ref{ququ}
associated with the operator $U$ and with
 ${\cal N}_1(\rho) \equiv \sigma_x
{\cal N}_0(\sigma_x \rho \sigma_x) \sigma_x$.
From  the properties of $\sigma_x$ it follows
that a Kraus set for ${\cal N}_1$ is given by the matrices~(\ref{kraus:qq}) by
exchanging $\theta$ and $\phi$. Consequently the
 Green function of this channel is given by
$G(\zeta,\xi)|_{\theta \leftrightarrow \phi}$
with $G(\zeta,\xi)$ as in Eq.~(\ref{gfqubitpure}).
Using this fact and the
linear dependence of Eq.~(\ref{GREEN}) with respect to ${\cal N}$
we can now  evaluate the Green function of the map~(\ref{eeew})
as follows
\begin{eqnarray}
G(\zeta,\xi) &=& q \; \delta^{(2)} \left(\zeta - \xi \cos \theta \cos \phi +\xi^*
\sin \theta \sin
\phi \right) \nonumber \\
&&\quad \times \exp \left[\frac{\cos (2 \theta) - \cos
(2 \phi)}{4} \xi \xi^*
   \right] \nonumber\\
&+& (1-q) \; \delta^{(2)} \left(\zeta - \xi \cos \phi \cos \theta +\xi^*
\sin \phi \sin
\theta \right) \nonumber \\
&&\quad \times \exp \left[\frac{\cos (2 \phi) - \cos
(2 \theta)}{4} \xi \xi^*
   \right] \;.
\label{orrenda}
\end{eqnarray}
Equation~(\ref{orrenda}) can finally be casted into the
form~(\ref{gfqubitmixed}) thanks to the identity
\begin{eqnarray}
q \; e^ {x \; \xi \xi^*}
+  (1-q) \; e^{-x \; \xi \xi^*}
&=&1 + (2q-1)\; x\; \xi \xi^*  \nonumber \\
&=& e^ {(2q-1) \; x \; \xi \xi^*}
\;,
\end{eqnarray}
which holds for all $x$ complex --- see Eq.~(\ref{fexp}). The
above is an example of a convex combination of Gaussian channels
(i.e., ${\cal N}_0$ and ${\cal N}_1$) which is still Gaussian.

A natural question is then whether or not the weakly complementary
channel~(\ref{duedue}) associated with Eq.~(\ref{eeew}) is also
Gaussian. To see this we first use the linearity of trace to
express the complementary  $\tilde{\cal N}$ as  a convex
combination of the weakly complementaries of ${\cal N}_0$ and
${\cal N}_1$, i.e., $\tilde{\cal{N}} = q \; \tilde{\cal N}_0 +
(1-q)\; \tilde{\cal N}_1$. Then we invoke the linearity of Eq.
(\ref{GREEN}) and use Eq.~(\ref{compl}) to write
 \bea
\tilde{G}(\zeta,\xi) &=& q\; {\delta^{(2)}} \left(\zeta - \xi \cos
\theta \sin \phi +\xi^* \sin \theta \cos
\phi \right) \nonumber \\
&& \quad \times \exp\left[ \frac{\cos (2 \theta) + \cos (2
\phi)}{4} \xi \xi^* \right] \nonumber \\
&+& (1-q) \ {\delta^{(2)}} \left(\zeta + \xi \sin \phi \cos \theta
-\xi^*  \cos \phi \sin \theta \right) \nonumber \\
&& \quad \times \exp\left[- \frac{\cos (2 \phi)+ \cos (2 \theta)
}{4} \xi \xi^* \label{gfmqubit1}
   \right] \;.
 \eea
This is of the form~(\ref{gfqubitmixed}) only for $q=0,1$.
Therefore, in general, the weakly complementaries of qubit-qubit
maps with mixed environment are not Gaussian even though they can
be expressed as a convex combination of Gaussian channels (i.e.,
$\tilde{\cal N}_0$ and $\tilde{\cal N}_1$). This can be pushed a
little further by observing that for generic choices of $\theta$,
$\phi$ and $q$, the weakly complementaries~(\ref{gfmqubit1}) are
not even unitarily equivalent to a qubit Gaussian
channel~\cite{NOTA10}.

\subsubsection{Weak-degradability properties}

Let us analyze the weak-degradability properties of the
qubit-qubit channels with mixed environment.

As in Sec.~\ref{ququ}  we prove that the maps ${\cal N}$ of
Eq.~(\ref{gfqubitmixed}) are weakly degradable for
$\cos(2\theta)/\cos(2\phi) \geqslant 0$. In this regime in fact
one can easily check that Eq.~(\ref{degrad}) can still be solved
with $G_x(\zeta,\xi)$ of the form~(\ref{gfmqubit1}) replacing
$\theta$ and $\phi$ with  $-\theta_x$ and $\phi_x+\pi/2$ where
$\theta_x, \phi_x$ satisfy the relations~(\ref{condiz2}).

Proving anti-degradability for  $\cos(2\theta)/\cos(2\phi)
\leqslant 0$ is not simple because, in general, $\tilde{\cal  N}$
is not in a Gaussian form --- see Eq.~(\ref{gfmqubit1}). However,
in this case we show that these channels cannot be used to
transfer quantum information since their quantum capacity
$Q$~\cite{QCAP} is null. To see this we notice that for
$\cos(2\theta)/\cos(2\phi) \leqslant 0$, ${\cal  N}$ is a mixture
(\ref{eeew}) of two channels (i.e., ${\cal N}_0$ and ${\cal N}_1$)
which are both anti-degradable and have hence null quantum
capacity, i.e., $Q({\cal N}_0)=Q({\cal N}_1)=0$
 --- see Appendix~\ref{s:degapp}.
Under these conditions it is easy to verify that also ${\cal N}$
must have a null $Q$. Indeed let us consider a new CPT map,
\begin{eqnarray}
{\cal  N}'(\rho) = q \ {\cal  N}_0 (\rho) \otimes
|0 \rangle_B \langle 0 | + (1-q) \ {\cal  N}_1 (\rho) \otimes|1
\rangle_B \langle 1 |\;, \nonumber
\end{eqnarray}
where $B$ is an ancillary system. We can now verify that the
${\cal N}$ is isomorphic to ${\cal E} \circ {\cal  N}'$ with
${\cal E}(...)=\mbox{Tr}_B[...]\otimes |0\rangle_B\langle 0|$
being a CPT map which replaces all states of $B$ with a fix given
output $|0\rangle_B$. Expressing $Q$ in terms of the output
coherent information~\cite{SCHUM} of the channel and using  the
quantum data processing inequality~\cite{nielsen} we can verify
that $Q({\cal  N}) \leqslant Q({\cal  N}')$. Besides, by using the
basic properties of von Neumann entropy~\cite{nielsen} we can
express the coherent information of ${\cal N}^\prime$ as $J({\cal
N}',\rho)=q J({\cal  N}_0,\rho) +(1-q) J({\cal  N}_1,\rho)$.
Putting all this together we get
\begin{eqnarray}
Q({\cal N}') &=& \lim_{N\to\infty} \max_{\rho}
J([{\cal N}']^{\otimes N},\rho)/N\nonumber \\
&\leqslant& q Q({\cal  N}_0) + (1-q) Q({\cal  N}_1) = 0
\;,
\end{eqnarray}
and hence $Q({\cal N})=0$.

\section{Conclusions}\label{s:conclusion}

In this work we introduce a characteristic function formalism for
the qubit channels in terms of generalized displacement operators
and Grassmann variables, inspired by a parallelism among these
maps and the Bosonic Gaussian channels.

We then present a Green function representation of the quantum
evolution that allows us to define the set of qubit Gaussian maps.
In this context, we find that all the Gaussian channels are
qubit-qubit, i.e., they can always be described in terms of a
unitary interaction of a qubit system with a {\em single} (not
necessarily pure) qubit environment. Similarly, it is known that
in the Bosonic case (almost) all the one-mode Bosonic Gaussian
maps are describable in terms of a single mode environment.

This formalism turns out to be elegant and powerful and, in
particular, it can be used to study the weak-degradability
properties of the qubit-qubit maps, for both pure and mixed qubit
environments, in terms of Green functions.

On one hand, in the case of pure environment, the qubit-qubit maps
are either degradable (i.e., additive coherent information) or
anti-degradable (i.e., Q=0). Besides, the complementary maps are
still qubit-qubit channels and so Gaussian. It is interesting to
note that an equivalent property holds for one-mode Bosonic
Gaussian channels. On the other hand, in the case of mixed
environment, we show that the qubit-qubit maps are either weakly
degradable or they cannot be used to transfer quantum information
(i.e., Q=0). However, in this case the weakly complementary maps
do not belong to the set of qubit-qubit channels and are not
Gaussian.

It is important to stress that this Green function formalism shows
clearly that the qubit Gaussian maps share analogous properties
with their continuous variable counterpart, i.e., the Bosonic
Gaussian channels.

Finally, we remark that the characteristic function approach,
introduced in this paper for qubit systems, can be generalized to
$d$-level quantum systems (qudit) in terms of generalized
Grassmann variables \cite{paragrass}.

\acknowledgments

This work was supported in part by the Centro di Ricerca Ennio De
Giorgi of the Scuola Normale Superiore of Pisa.

\appendix

\section{Grassmann variables} \label{app1}

A Grassmann variable $\xi$ spans over a set of objects (the
Grassmann numbers) $\xi_1$, $\xi_2$, $\cdots$, which anti-commute.
Indeed, given any $\xi_i$ and $\xi_j$ elements of the set, they
satisfy the relation \bea \xi_i \xi_j   =  - \xi_j\xi_i\;,
\label{ccc} \eea while obeying ordinary commutation relations with
respect to the multiplication by a complex number. In particular
Eq.~(\ref{ccc}) implies that a Grassmann variable is
$2$-nilpotent, i.e., $\xi^2 = 0$ (note that $0$ is trivially
included in the Grassmann variable set). At a mathematical level,
the above conditions can be rigorously formalized by saying that
Grassmann numbers are the generators of an algebra over the
complex field which obey anti-commutation relations.

Complex conjugation of $\xi$ can be defined by introducing an
extra Grassmann variable $\xi^*$ whose elements $\xi_1^*$,
$\xi_2^*$, $\cdots$ obey the same relation~(\ref{ccc}) and
anti-commute with all the $\xi_i$s, i.e., \bea
\xi_i^* \xi_j^*   &=&  - \xi_j^*\xi_i^*\;, \\
\xi_i^* \xi_j   &=&  - \xi_j\xi_i^*\;. \label{ccc1} \eea To
identify $\xi_i^*$ with the complex conjugate of $\xi_i$ we
finally require the relations
 \bea
(\xi_i^*)^* &=& \xi_i \nonumber \\
(\xi_i x)^* &=& x^* \xi_i^* \label{complexconj} \eea to be
satisfied for any $x$ complex number or product of the $\xi_1,
\xi_2, \cdots $ and $\xi_1^*, \xi_2^*,\cdots$.

Given the above properties it follows that the most general
function $ f( \xi ,\xi^*) $  is linear both in  $ \xi $ and $
\xi^* $, i.e.,
 \beq
 f( \xi ,\xi^*) = A + B_1 \xi + B_2 \xi^* + C \xi^* \xi \;,
\label{fx}
 \eeq
with $A$, $B_{1,2}$, and $C$ independent from $\xi$ and $\xi^*$.
In particular, the exponentials become \bea &&\exp( B_1\xi  + B_2
\xi^* + C \xi^* \xi) \equiv
\sum_{n=0}^\infty\frac{( B_1\xi  + B_2 \xi^* + C \xi^* \xi)^n}{n!} \nonumber \\
&& = 1   + B_1 \xi + B_2 \xi^*  + C \xi^* \xi + B_1 \xi B_2
\xi^*/2 +  B_2 \xi^* B_1 \xi /2 \;. \nonumber \\ \label{fexp} \eea
This expression can be used to verify that (apart from a global
multiplicative term) {\em any} function~(\ref{fx}) can be written
as an exponential.

Integration over $\xi$ and $\xi^*$ can be defined by introducing
the ``differential'' $d \xi$ and $d\xi^*$. These are assumed to
obey  the same anti-commutation relations obeyed by the variables
$\xi$ and $\xi^*$, including Eqs.~(\ref{ccc}),~(\ref{commop}), and
(\ref{vec}). The integrals are then defined according to the
Berezin rules \bea
\int \! d\xi & = & \int \! d\xi^* = 0 \;, \nonumber \\
\int \! d\xi \, \xi & = & \int \! d\xi^* \, \xi^* =1 \;.
\label{integral} \eea Joint integration with respect to $\xi$ and
$\xi^*$ is finally defined by identifying the double differential
$d^2 \xi$ as follows, \beq d^2\xi \equiv d\xi^* \, d\xi = -d\xi \,
d \xi^* \;. \eeq In this context one can identify an analogous of
the Dirac delta function $\delta^{(2)}(\mu -\nu)$ in the complex
plane. Such Grassmann delta is defined as
 \bea
 \delta^{(2)} ( \xi - \zeta)  & \equiv &
\int \! d^2 \kappa  \, \exp \left[
 \kappa \left( \xi^* - \zeta^* \right)
- \left( \xi - \zeta \right) \kappa^* \right]  \nonumber \\
& = & \left( \xi - \zeta  \right) \, \left(
\xi^*  - \zeta^* \right)\;, \label{deltaf}
 \eea
with $\xi$, $\zeta$, and $\kappa$ Grassmann variables. Indeed,
from Eq.~(\ref{integral}) and from Eq.~(\ref{fx}) we have \beq
\int \! d^2 \xi \; \delta^{(2)}( \xi - \zeta ) \, f ( \xi, \xi^* )
= f (\zeta,\zeta^*) \;,
 \eeq
for all $ f ( \xi, \xi^* )$. Notice that the delta
function~(\ref{deltaf}) commutes with any Grassmann numbers and
satisfies the relation $ \delta^{(2)} ( \xi - \zeta ) =
\delta^{(2)} (\zeta - \xi ) = - \delta^{(2)} (\xi^* - \zeta^*)$.

A useful property is the following. Given the function
$f(\xi,\xi^*)$ one can define its even and odd parts, i.e., \bea
f_{\pm}(\xi,\xi^*) \equiv \frac{f(\xi,\xi^*) \pm
f(-\xi,-\xi^*)}{2} \; . \label{evenodd} \eea According to
Eq.~(\ref{fx}) they are of the form $f_+(\xi,\xi^*) = A + C \xi^*
\xi $ and $f_-(\xi,\xi^*)  = B_1 \; \xi  + B_2 \; \xi^*$,
respectively. Now given $g(\xi,\xi^*)$ another function we can
write \bea \int d^2 \xi \; f_{\pm} (\xi,\xi^*) g_{\mp}(\xi,\xi^*)
&=&  0 \nonumber \eea and thus \bea &&\int d^2 \xi \; f
(\xi,\xi^*) g(\xi,\xi^*) =
\int d^2 \xi \; f_{+}(\xi,\xi^*) g_{+} (\xi,\xi^*) \nonumber \\
&&\qquad\qquad + \int d^2 \xi \; f_{-}(\xi,\xi^*) g_{-}
(\xi,\xi^*) \;. \label{useful} \eea

\subsection{More about trace} \label{app2}

Equation~(\ref{traccia}) shows that the cyclicity of the trace
needs to be modified when involving Grassmann terms. If we need to
move only qubit operators, then the standard  rule applies, i.e.,
\bea \Tr [ \Theta_1 \xi_1  \cdots \Theta_n \xi_n \Theta_{n+1} ]
 &=&  \Tr[ \Theta_{n+1} \Theta_1 \xi_1 \cdots
\Theta_n \xi_n  ] \nonumber \\
 &=&  \Tr[\xi_1
\cdots \Theta_n \xi_n   \Theta_{n+1} \Theta_1  ] \nonumber
 \;. \\
\label{QUIE} \eea
 On the contrary, if we move also Grassmann
variables, by exploiting the anti-commutation rules of the
$\xi_i$s, we get \bea &&\Tr [ \Theta_1 \xi_1 \Theta_2 \xi_2 \cdots
\Theta_n \xi_n
\Theta_{n+1} ] \\
 && \quad \qquad = (-1)^{n-1} \Tr[ \xi_n \Theta_{n+1} \Theta_1 \xi_1
\cdots \Theta_n ] \nonumber \\
&& \quad \qquad = (-1)^{n-1} \Tr[ \Theta_2  \xi_2 \cdots \Theta_n
\xi_n \Theta_{n+1} \Theta_1  \xi_1] \ . \nonumber
 \eea
Finally in conjunction with Eq.~(\ref{adjont}),
Eq.~(\ref{traccia}) gives \bea &&\Big(\Tr [ \Theta_1 \xi_1
\Theta_2 \xi_2 \cdots \Theta_n \xi_n \Theta_{n+1} ]
\Big)^* \label{tracciaag} \\
&& \qquad \qquad = \Tr[\Theta_{n+1}^\dag \; \xi_n^* \Theta_n^\dag
\cdots \xi_2^*  \; \Theta_2^\dag \; \xi_1^* \;
\Theta_1^\dag] \;.\nonumber \eea

\section{Weak-degradability vs. Anti-degradability}\label{s:degapp}

It is a well known (see, e.g., \cite{HPPI}, \cite{LINDBLAD}) that
any CPT map ${\cal N}$ can be described by  a unitary coupling
between the system $S$  with an external ancillary system $E$
(describing the {\em environment}) prepared in some fixed {\em
pure} state. This follows from the Stinespring
dilation~\cite{STINE} of the map which is unique up to a partial
isometry. More generally, one can describe ${\cal N}$ as a
coupling with an environment prepared in some {\em mixed} state
$\rho_E$, i.e.,
 \begin{eqnarray}
{\cal N}(\rho) = \mbox{Tr}_E[ U(\rho \otimes \rho_E) U^\dag] \;,
\label{HGCuno}
\end{eqnarray}
where $\mbox{Tr}_E [ ... ]$ is the partial trace over the
environment $E$ and $U$ is a unitary operator in the composite
Hilbert space ${\cal H}_S \otimes {\cal H}_E$. As proposed in
Ref.~\cite{NOSTRO1} we call Eq.~(\ref{HGCuno}) a ``physical
representation'' of ${\cal N}$  to distinguish it from the
Stinespring dilation, and to stress its connection with the
physical picture of the noisy evolution represented by ${\cal N}$.
Moreover, Eq.~(\ref{HGCuno}) motivates the following
definition~\cite{NOSTRO,NOSTRO1}. For any physical representation
in Eq.~(\ref{HGCuno}) of the quantum channel ${\cal N}$ we define
its {\em weakly complementary} as the map ${\tilde{\cal N}}$ which
takes the input state $\rho$ into the state of the environment $E$
after the interaction with $S$, i.e.,
\begin{eqnarray}
\tilde{\cal N}(\rho) = \mbox{Tr}_S[ U (\rho \otimes \rho_E)
U^\dag] \;. \label{duedue}
\end{eqnarray}
The transformation~(\ref{duedue}) is CPT and describes a quantum
channel connecting systems $S$ and $E$. It is a generalization of
the {\em complementary channel} ${\cal N}_{\text{com}}$ defined in
Refs.~\cite{DEVSHOR,HOLEVOREP,KING}. If some channel ${\cal T}$
does exist such that
\begin{eqnarray}
({\cal  T}\circ {\cal N})(\rho) = \tilde{\cal N}(\rho) \;, \label{degrado}
\end{eqnarray}
for all density matrices $\rho$, then ${\cal N}$ is called  {\em
weakly degradable} and $\tilde{\cal N}$ {\em anti-degradable}.
Similarly if
\begin{eqnarray}
(\overline{\cal T}\circ \tilde{\cal N})(\rho) = {\cal N}(\rho) \;,
\label{antidoto}
\end{eqnarray}
for  some channel $ \overline{\cal T}$ and all density matrices
$\rho$, then  $\cal N$ is {\em anti-degradable} while $\tilde{\cal
N}$ is {\em weakly degradable} (see~\cite{NOSTRO,NOSTRO1}). In
Ref.~\cite{DEVSHOR} the channel $\cal N$ is called {\em
degradable} if one considers the environment in a pure state.
Clearly any degradable channel~\cite{DEVSHOR} is weakly degradable
but the opposite is not necessarily true.

Degradability and anti-degradability have been proved useful to
analyze the quantum capacity~\cite{QCAP} of the channel. On one
hand, one can verify that anti-degradable channels (where this
property is defined irrespectively from the purity of $\rho_E$
associated with the physical representation) cannot be used to
convey quantum messages in reliable fashion --- i.e., their
quantum capacity $Q$ nullifies~\cite{QC,NOSTRO,NOSTRO1} . On the
other hand, instead degradable channels~\cite{DEVSHOR} allows for
a single letter formula expression for $Q$ --- i.e., the maximum
of their output coherent information is additive.


\begin{thebibliography}{99}
\bibitem{NOTAG}
The characteristic function of the state~(\ref{rho}) can be
written as $\chi(\xi) \equiv \exp[ \gamma \xi - \gamma^* \xi^* +
(2p-1)\xi \xi^*/2]$.
\bibitem{NOTA1}
The fact that $c$ must be real can be derived by imposing the
Hermitianity constraint Eq.~(\ref{hermi}) to the output
characteristic function~(\ref{goal}).
\bibitem{NOTA10}
It is worth observing that in the canonical form~(\ref{canono})
the weakly complementaries~(\ref{gfmqubit1}) are characterized by
$\lambda_1=(2 q-1)\sin(\theta+\phi)$,
$\lambda_2=(2q-1)\sin(\phi-\theta)$,
$\lambda_3=[\cos(2\theta)-\cos(2\phi)]/2$, $t_1= t_2=0$ and $t_3=
(2 q-1)/2[\cos(2\theta) + \cos(2\phi)]/2$. Since $\lambda_1
\lambda_2 \neq \lambda_3$, this is an indirect way of verifying
that these maps are not Gaussian~(\ref{gfqubitmixed}). However,
the canonical form~\cite{ruskai2} is  uniquely determined only up
to unitary transformations acting on the input and on the output
of the map. Applying such unitary transformations one can indeed
permute the  $\lambda$s. After such permutations, one can have,
for instance, $\lambda_1=[\cos(2\theta)-\cos(2\phi)]/2$,
$\lambda_2=(2 q-1)\sin(\theta+\phi)$, $\lambda_3=(2
q-1)\sin(\phi-\theta)$. Now $\lambda_1 \lambda_2 = \lambda_3$ can
be satisfied for some particular values of $\theta$ and $\phi$,
i.e., $\theta=\phi$, $\theta+\phi=\pi/2, \ 3\pi/2, \ 5\pi/2, \
7\pi/2$, and $\theta=\phi \pm \pi$. In these cases we can say that
the weakly complementaries~(\ref{gfmqubit1}) are unitarily
equivalent to a Gaussian channel.
\end{thebibliography}
\end{document}